\documentclass[12pt]{iopart}

\usepackage{cite}
\usepackage{iopams}
\usepackage{amssymb}
\usepackage{amscd}
\usepackage{wasysym}
\usepackage{color}
\usepackage{dsfont}

\usepackage[utf8]{inputenc}
\usepackage[pdftex]{graphicx}
\DeclareGraphicsExtensions{.pdf}

\newcommand{\bq}{\begin{equation}}
\newcommand{\eq}{\end{equation}}
\newcommand{\ba}{\begin{eqnarray}}
\newcommand{\ea}{\end{eqnarray}}
\newcommand{\be}{\begin{equation}}
\newcommand{\ee}{\end{equation}}
\newcommand{\bea}{\begin{eqnarray}}
\newcommand{\eea}{\end{eqnarray}}

\begin{document}

\bibliographystyle{iop}

\title{Curing critical links in oscillator networks as power grid models}

\author{Martin Rohden${}^{1}$, 
           Dirk Witthaut${}^{2,3,4}$, 
           Marc Timme${}^{4,5}$, and
           Hildegard Meyer-Ortmanns${}^{1}$
           }

\address{${}^1$Jacobs University Bremen, Physics and Earth Sciences, 28759 Bremen, Germany \\
             ${}^2$Forschungszentrum J\"ulich, Institute for Energy and Climate Research - 
                Systems Analysis and Technology Evaluation (IEK-STE), 52428 J\"ulich, Germany \\
             ${}^3$Institute for theoretical Physics, University of Cologne, 50937 K\"oln, Germany \\
             ${}^4$Network Dynamics, Max Planck Institute for Dynamics and Self-Organization (MPIDS),
                  37077 G\"ottingen, Germany\\
             ${}^5$Faculty of Physics, Georg August University G\"ottingen, Germany
}

\ead{m.rohden@jacobs-university.de}

\date{\today }

\begin{abstract}
Modern societies crucially depend on the robust supply with electric energy. Blackouts of power grids can thus have far reaching consequences. During a blackout, often the failure of a single infrastructure, such as a critical transmission line, results in several subsequent failures that spread across large parts of the network. Preventing such large-scale outages is thus key for assuring a reliable power supply. Here we present a non-local curing strategy for oscillatory power grid networks based on the global collective redistribution of loads. We first identify critical links and compute residual capacities on alternative paths on the remaining network from the original flows. For each critical link, we upgrade lines that constitute bottlenecks on such paths. We demonstrate the viability of this strategy for random ensembles of network topologies as well as topologies derived from real transmission grids and compare the nonlocal strategy against local back-ups of critical links. These strategies are independent of the detailed grid dynamics and combined may serve as an effective guideline to reduce outages in power grid networks by intentionally strengthen optimally selected links.
\end{abstract}


\submitto{\NJP}

\maketitle

% --- content ------------------------------------------------

\section{Introduction}

We are currently witnessing a rapid transition in power generation from conventional to renewable power sources. Typically, renewable power sources are strongly fluctuating, have a lower power output than conventional ones and their potential geographical locations are restricted to places with sufficient solar or wind power \cite{Turn99,Sims11}. This development is challenging the operation and stability of electric power grids: Power has to be transmitted over large distances \cite{Pesc14}, Fluctuations must be balanced by storage or backup power plants \cite{Heid10,Rasm12,Mila13} and many decentralized units must be controlled \cite{Amin05,Rohden12,Scha15}.  

In general power grids work reliably, and power outages on a large scale are rare events \cite{Fair04,Pour06,Hine09}. However, just in the last decade major outages were recorded in India (2012), Bangladesh (2014), Pakistan (2015), Indonesia (2005), Brazil (2009), Turkey (2015) and Germany (2006). A detailed analysis of the latter outage can be found in \cite{Report2006}. Each of these outages affected millions of people with potentially destructive impact \cite{Helb13,Brum13}. It is commonly expected that the loads will increase strongly in future grids \cite{Pesc14}, so that these outages can become more likely. So it is urgent to trace back what causes such outages and to minimize the risk for such events to happen in the future. It may come as a surprise that large power outages in the past often were triggered by the outage of a single infrastructure, such as a transmission line \cite{Pour06}. In such an event, after one transmission line failed, a second line became overloaded and initiated a whole cascade of further failures \cite{Albe00,Albe04,Simo08,Buld10,Schn11,12braess,13nonlocal}.

To exclude such outages induced by single infrastructure failure, the so-called ($N-1$) rule  is currently implemented in modern power grid
operations. The ($N-1$)-rule states that at every instant of time, the power grid has still to be fully functional even if any given single
infrastructure, e.g. a transmission line, fails \cite{Prab94}. However, this criterion may be violated in times of high loads, as for
example in Germany (2006). The intentional shutdown of a single transmission line in Northern Germany brought the entire grid to the edge of a
breakdown. A coupling of the busbars at a nearby transformer station then finally triggered the blackout, in which the European grid
fragmented into three mutually asynchronous areas \cite{Report2006}. Additionally, an inclusion of new power sources into the grid is ongoing, and the need
for additional transmission lines continues. Therefore it becomes more and more challenging to maintain the implementation of the ($N-1$)-rule.

In this article we propose a nonlocal strategy to cure networks from critical line failures. The strategy is based on a graph-theoretical quantification of the redundant flows along paths of a supply network. Our analysis reveals that bottlenecks along those paths limit the stable operation of the grid. The main cure then comes from improving the capacities of the lines that constitute these bottlenecks. We compare this nonlocal strategy to the much simpler, local strategy of building local backup transmission lines in parallel to critical lines, thereby taking over the power flow if a critical line fails. We demonstrate that for example networks such as the Romanian grid topology (Fig. 1), an increase of only the bottlenecks' capacities is an efficient countermeasure against outages, less costly than building additional transmission lines in parallel to critical ones. The non-local strategy is successful by exploiting unused 'resource' capacities the network exhibits as a whole. It ceases to work if no such resources exist, as in some types of random networks.

\section{An oscillator model for power grid dynamics}

\begin{figure}[tb]
\centering
\includegraphics[height=5cm]{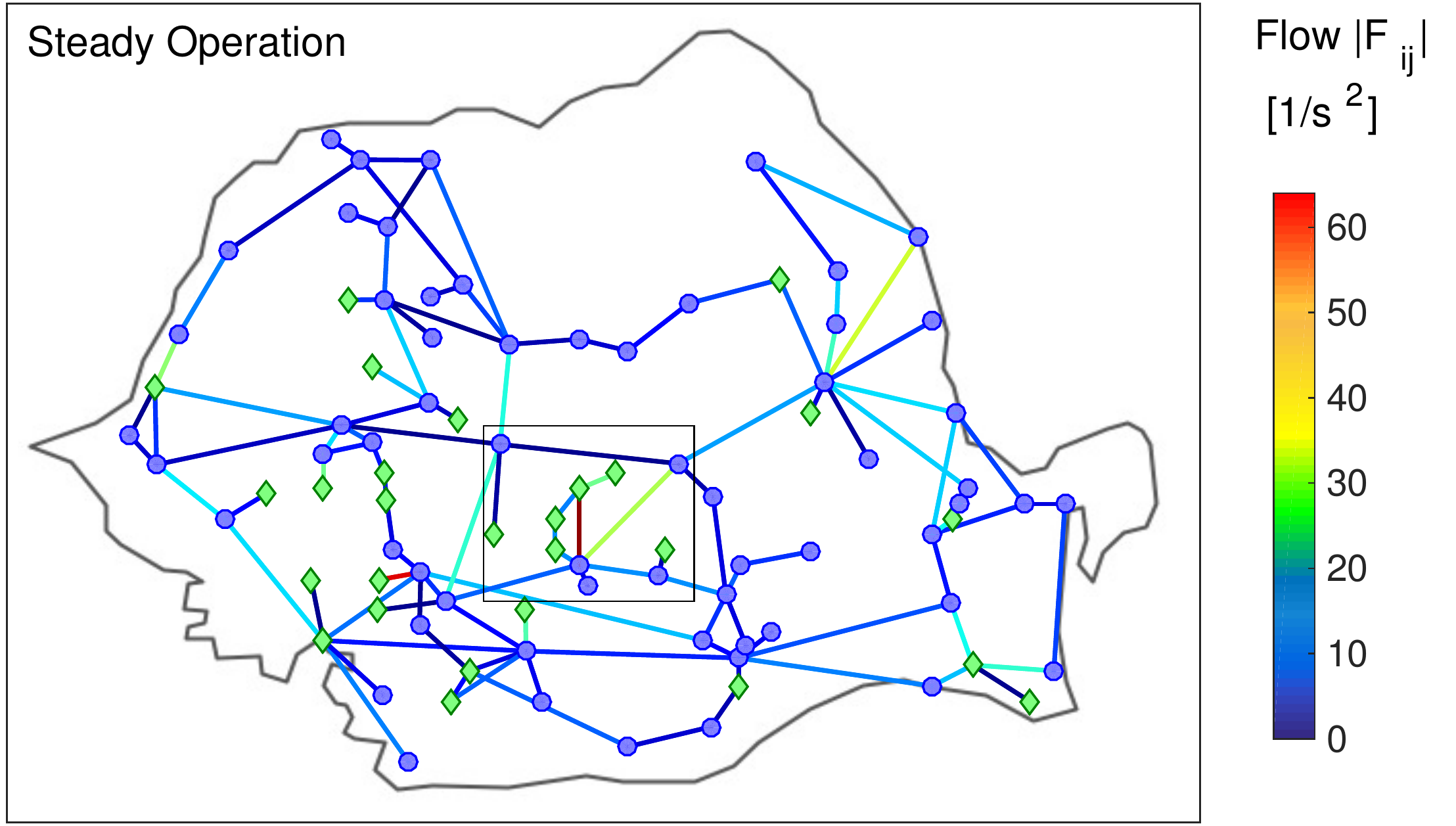}
\caption{\label{fig:figure1} (Color online) Test grid based on the topology of the Romanian high-voltage power grid according to \cite{rogrid} of 88 nodes, out of which 22 are generators
(blue circles) and 66 are consumers (green diamonds). The links correspond to transmission lines. The actual load of the links in a steady state is
coded by the colors (blue representing low load, red high load). A magnification of the marked area is illustrated in Fig.~\ref{fig:figure1b}. For further information see the text.
}
\end{figure}

In this article we focus on the rotor angle and frequency stability of AC power grids with respect to transmission line outages. We model the dynamics of the power grid as a network of $N$ rotating synchronous machines representing electric motors or generators \cite{Berg81,Fila08,Rohden12,Rohden14,Menc14}, as they are used in hydro-electric power plants and the like. Each machine $j \in \{1,\ldots,N\}$ is characterized by the mechanical power $P^{\rm mech}_j$ it generates, ($P^{\rm mech}_j > 0$) for a synchronous generator, or it consumes ($P^{\rm mech}_j < 0$) for a synchronous motor. The state of each rotating machine $j$ at time $t$ is given by its mechanical phase angle $\theta_j(t)$ and its velocity $d \theta_j / dt$.

The dynamics of the $j$th oscillator is then determined by the equation
\begin{equation}
   M_j \frac{d^2 \theta_j}{dt^2} + D_j \frac{d \theta_j}{dt}
    = P_j^{\rm mech}  - P_j^{\rm el},
   \label{eqn:eom-theta}
\end{equation}
where $P_j^{\rm el}$ is the electric power exchanged with the grid and $P_j^{\rm{mech}}$ is the mechanical power, related to the turbine torque $\tau_t$ via $\tau_t\cdot \omega_0$, $\omega_0$ being the base frequency of the grid. $M_j$ is the machine's angular mass, $D_j$ the damping coefficient of the torque.
We neglect ohmic losses  such that the admittance $Y_{jk}$ of a transmission line between nodes $j$ and $k$ is
purely imaginary $Y_{jk}=i\;B_{jk}$. Assuming furthermore that the magnitude of the voltage is constant throughout the grid, $\vert V_j \vert=V_0$ for all nodes $j \in \{1,\ldots,N\}$ , the electric power flow from node $j$ to node $k$ is given by
\begin{equation}\label{eq2}
    F_{jk} =  V_0^2 B_{jk}  \sin(\theta_j - \theta_k),
\end{equation}
and the total electric power acting on node $j$ is obtained by summing over all flows to the connected nodes, $P_j^{\rm el} = \sum_{k=1}^N F_{kj}$.

During regular operation, generators as well as consumers run with the same base frequency within the grid:
$\omega_0 = 2 \pi \times 50 \, {\rm s}^{-1}$  or $\omega_0 = 2 \pi \times 60 \, {\rm s}^{-1}$, respectively. The phase of each element is
then written as
\begin{equation}
	\theta_j(t) = \omega_0 t + \phi_j(t),
	\label{eqn:phase}
\end{equation}	
where $\phi_j$ denotes the phase difference to the reference phase $\omega_0 t$. If we insert (\ref{eqn:phase}) into (\ref{eqn:eom-theta}), we obtain the swing equation for the phase deviations $\phi_j(t)$ with $P^m=P^{\rm{mech}}-D\omega_0$ and $P^m$ the net shaft power input there. Using furthermore  (\ref{eq2}), we obtain  the grid dynamics, which
we study in this article and call ``oscillator model" for brevity in the following:
\begin{equation}
   \frac{d^2 \phi_j}{dt^2} = P_j - \alpha_j \frac{d \phi_j}{dt}
          + \sum_{k=1}^N K_{jk}\sin(\phi_k - \phi_j) ,
        \label{eq.7}
\end{equation}
where $P_j\equiv \frac{P^m_{j}}{M_j} =\frac{P_j^{\rm mech} - D_j \omega_0}{M_j}$, $\alpha_j=\frac{D_j}{M_j}$, $K_{jk}=\frac{V_0^2 B_{jk}}{M_j}$.
In this formulation the regular operation of the grid corresponds to a \emph{steady state} with $d\phi_j/dt = 0$ for all nodes $j$, compare
\cite{Timm07,Dorf13,Mott13}. The network topology is implemented via the capacity matrix $K_{ij}$. The networks we consider are undirected,
$K_{ij}= K_{ji}$, $K_{ij}=0$ if node $i$ is not connected with $j$, while the flows $F_{ij}$, assigned to the links, are directed.

In the so-called structure preserving model \cite{Berg81,Dorf13} one assumes frequency-dependent loads which lead to the same equations of motion as in (\ref{eqn:eom-theta}) with $M_j=0$ for load nodes. The steady
state is the same as for the oscillator model such that our results also apply to the structure preserving model.
A comparison of different models for synchronous machine operation with and without voltage dynamics can be found in \cite{Weck13,Schmie14,Auer15,Nish15}.

\section{Curing strategies}
\label{sec:kred}

\begin{figure}[tb]
\centering
\includegraphics[height=5cm]{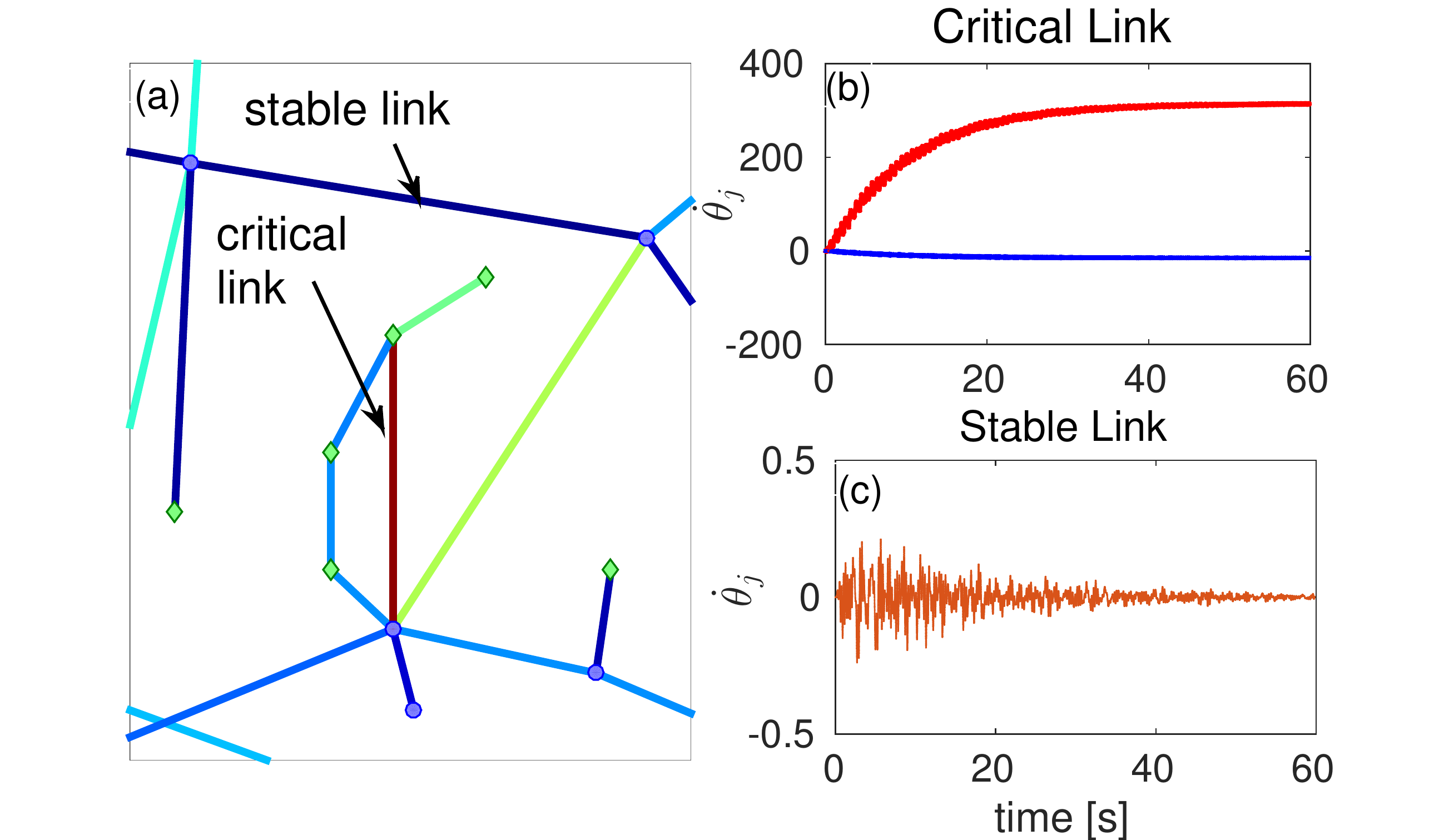}
\caption{\label{fig:figure1b}(Color online) Critical and stable links in the Romanian power grid. (a) Zoom into the grid structure in central Romania, with
one stable and one critical link marked by arrows. (b) Divergence of the common frequency $\dot{\theta}$ towards two clusters at high and low
oscillating frequencies after the marked critical link is removed. (c) Recovering of the net frequency with $\dot{\theta_i}=0$ for all nodes
if the marked stable link is removed.
}
\end{figure}

A failure of a single transmission line can have different types of impact on the operation of the power grid. It may cause only weak transient
disturbances such that the network relaxes back to a stable steady state, or it may destabilize the synchronous action of the machines and, in
the extreme case, induce a cascade of failures that ends up in a large-scale power outage. We call links, whose failure could induce power outages,
critical links. All other links are referred to as stable. It is thus desirable to identify and stabilize all potentially critical links \cite{16critical}.

First we identify critical links by removing single links from the network and recording the time evolution of all rotator phases and their velocities.
For stable links the phase differences become constant over time, and the common frequency is the desired net frequency. For critical links,
deviations from the net frequency outside a tolerance interval remain, which would lead to an emergency shutdown of parts of the power grid in
real-world power grids. An example is demonstrated in Fig.~\ref{fig:figure1b}, where a stable and a critical link are marked in panel (a). The
response of the frequency to a failure of the marked critical link is illustrated in panel (b) and to the marked stable link in panel (c).

We analyze two different strategies to stabilize critical links. A straightforward local strategy (later called strategy II) is to build backup links for the
critical links, i.e., to increase the grid's redundancy locally at the places, where critical failures take place.

Strategy I is in general non-local and works as follows. If the link $(a,b)$ fails, the (directed) flow $F_{ab}$ has to be rerouted over alternative paths in
the network. However, the links along these paths have only a limited residual capacity $K_{ij} - F_{ij}$ to take over this flow. Correspondingly,
we here determine the redundant capacity $K_{ab}^{\rm red}$ as the minimum value of the residual capacities along all shortest alternative paths $p$
(apart from the direct connection between a and b) connecting $a$ with $b$:
\begin{equation}
     K_{ab}^{\rm red} = \min_{(i,j) \in p}  (K_{ij}-F_{ij}).
    \label{eq:kred-tilde}
\end{equation}
The link $(r,s)$, which actually carries and determines the redundant capacity $K_{ab}^{\rm red}$ with respect to the removed link $(a,b)$, i.e.~the link for which
\begin{equation}
    K_{rs}-F_{rs} = \min_{(i,j) \in p}  (K_{ij}-F_{ij}),
    \label{eq:def-bottleneck}
\end{equation} 
represents the bottleneck for
the load redistribution. If no alternative shortest path exists, the link $(a,b)$ is a bridge and therefore a bottleneck. The minimum at the
link $(r,s)$ may be degenerate with several bottlenecks along the paths in question. Strategy I then amounts to increasing the capacity of each
of these bottlenecks by a factor $c>1$
\begin{equation}
   K_{rs}' = c \times K_{rs}
   \label{eqn-increaseK}
\end{equation}
rather than reinstalling the broken link $(a,b)$, once it turned out to be critical.

Schematically, we use the following  algorithm for strategy I: For a given uniformly chosen capacity $K_{ij}$ calculate the load distribution in
the normal operation of the grid. Sort the links in an arbitrary sequence. Start the loop over all links $(a,b)$:

\renewcommand{\theenumi}{\arabic{enumi}}
\begin{enumerate}
\item Determine whether link $(a,b)$ of the sequence is critical:
\begin{enumerate}
\item Eliminate $(a,b)$, i.e., set $K_{ab}=0$ and run the dynamics.
\item If the system approaches a stable state with $\dot{\theta_i}=0$ for all nodes $i$, go to the next link in the sequence without any update, otherwise the link is identified as critical.
\end{enumerate}

\item If the link $(a,b)$ is critical, find and strengthen the bottleneck:
\begin{enumerate}
\item Calculate the redundant capacity according to equation (\ref{eq:kred-tilde}) and identify the bottleneck link $(r,s)$ according to equation (\ref{eq:def-bottleneck}).
\item Increase the capacity at the bottleneck according to $c \times K_{rs}$. Here we choose $c=1.1$.
\item Determine whether the link is still critical despite the increased capacity $K'_{r,s}$. If no, go to the next link.
\item If yes, determine the new bottleneck with updated
capacity according to (\ref{eq:kred-tilde}), with $K_{rs}$ replaced by $c\cdot K_{rs}$, and increase its capacity again according to (\ref{eqn-increaseK}). Repeat these steps until the
link becomes stable.
\end{enumerate}

\item Calculate the sum of the increase in capacities along the bottlenecks of all critical links to decide which strategy is superior. If the
sum according to the here proposed strategy I needs less additional capacity than building backup lines with capacity $K_{ab}$ for each critical
link, strategy I is superior.
\end{enumerate}

\begin{figure}[tb]
\centering
\includegraphics[height=5cm]{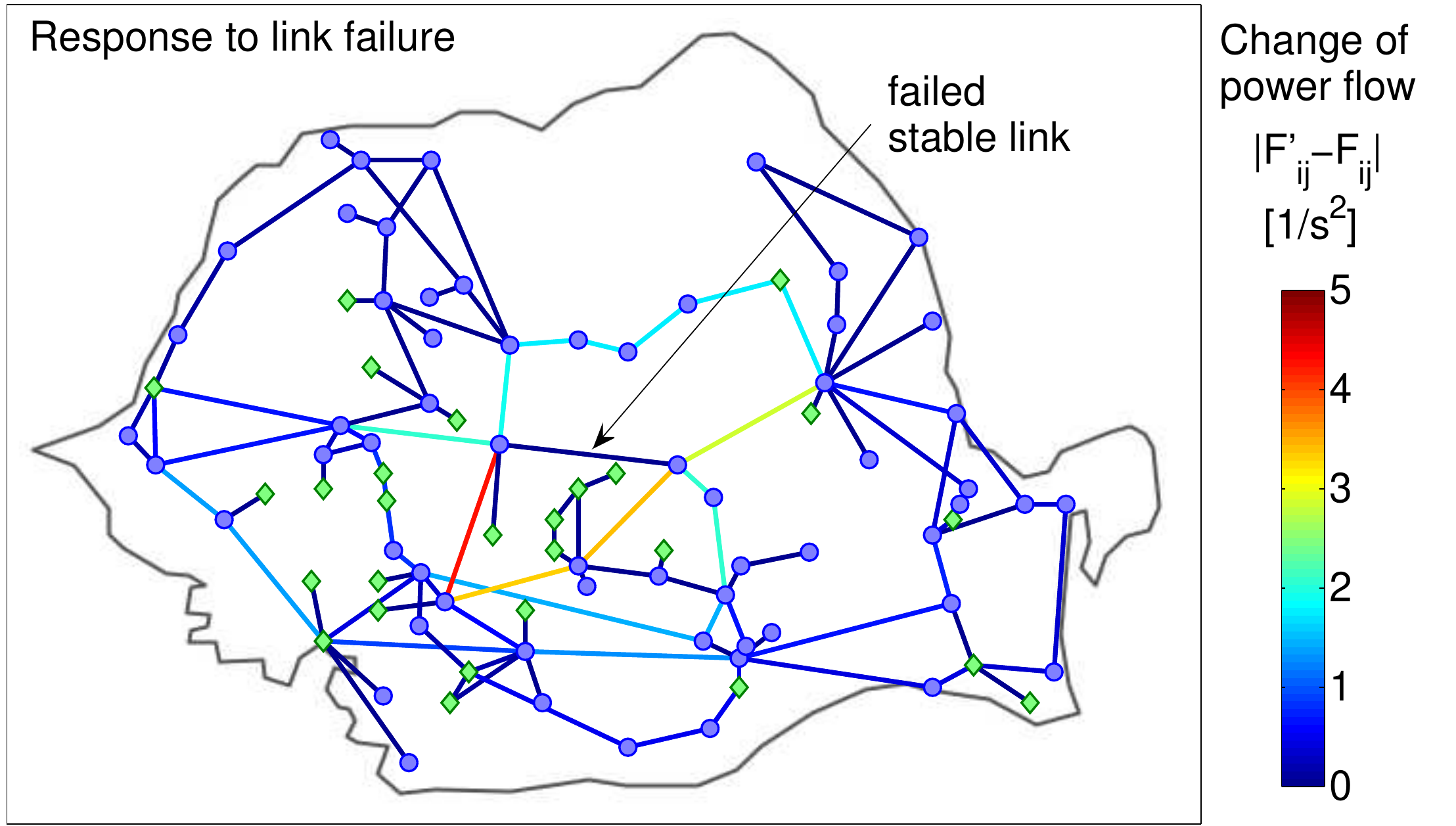}
\caption{\label{fig:figure4} (Color online) Response to the failure of a stable link. Fast drop off of the additional load with the distance from the removed link when the
load has to be redistributed after the deletion of a stable link. The color codes the difference between the loads $F^\prime-F$ after and before the removal.
It is pronounced only along the shortest alternative paths connecting the endpoints of the stable link.}
\end{figure}

Some remarks are in order. It should be noticed that $K_{ab}^{\rm red}$ is not the capacity of the link $(a,b)$ itself, but depends on $(a,b)$ via the shortest alternative paths. Furthermore, the flow $F_{ij}$ follows
from the original dynamics with the critical link $(a,b)$ being present. A redistributed flow $F^{\prime}_{ij}$ after the link's removal is only accessible if the removed link $(a,b)$ is stable,
but not if it is critical and the power flow is shut down. The rationale behind choosing the flow $F_{ij}$ before the removal is that the bottleneck determined from this original flow distribution is responsible for the instability when the link
$(a,b)$ is removed. Finally, the focus on the shortest out of all alternative paths assumes that the difference due to the redistributed load decays fast with the distance
from the removed critical link. Actually, in the Romanian grid we find that rerouting takes place predominantly over the shortest alternative paths.
Fig.~\ref{fig:figure4} illustrates the flow redistribution $|F_{ij}-F^{\prime}_{ij}|$ after the marked stable link is removed. The flow difference is most
pronounced along the shortest alternative paths. We tested the redistribution of flows for other stable links and found similar decay behavior with the
distance from the removed link. (Fast decay behavior is, of course, not automatically guaranteed, see e.g.\cite{darka}.)

Our strategy applies to transmission line overloads, but do not capture voltage instabilities. For models, dealing with transmission line overloads, they are generally independent of the specific model system. The specific dynamics enters in strategy I only where the bottlenecks are determined from the flow
redistribution. Other improved or refined models may replace the oscillator dynamics, as long as they satisfy the continuity equation for flows, and losses along the lines are negligible.

\section{Performance of curing strategies}

\subsection{Curing the Romanian power grid}

We test the performance of the strategies using a synthetic network, sharing the topology with the Romanian high voltage power transmission grid \cite{rogrid}  as
illustrated in Fig.~\ref{fig:figure1} and the distribution of generators and consumers including their actual values for production and consumption. However, we discard all connections to neighboring countries and keep only the internal lines of Romania. The net power at the
66 substations is negative with $P \in [-25,-2]\ {\rm s}^{-2}$, i.e., these nodes act as consumers (blue circles in Fig.~\ref{fig:figure1}). There are
22 substations with positive power and $P \in [0...110]\ {\rm s}^{-2}$ (green diamonds). For simplicity we assume that the transmission capacity of all
links is equal. Furthermore, we vary the value of $K$ to interpolate between a weakly ($K$ large) and a strongly ($K$ small) loaded grid.

\begin{figure}[tb]
\centering
\includegraphics[width=0.45\textwidth]{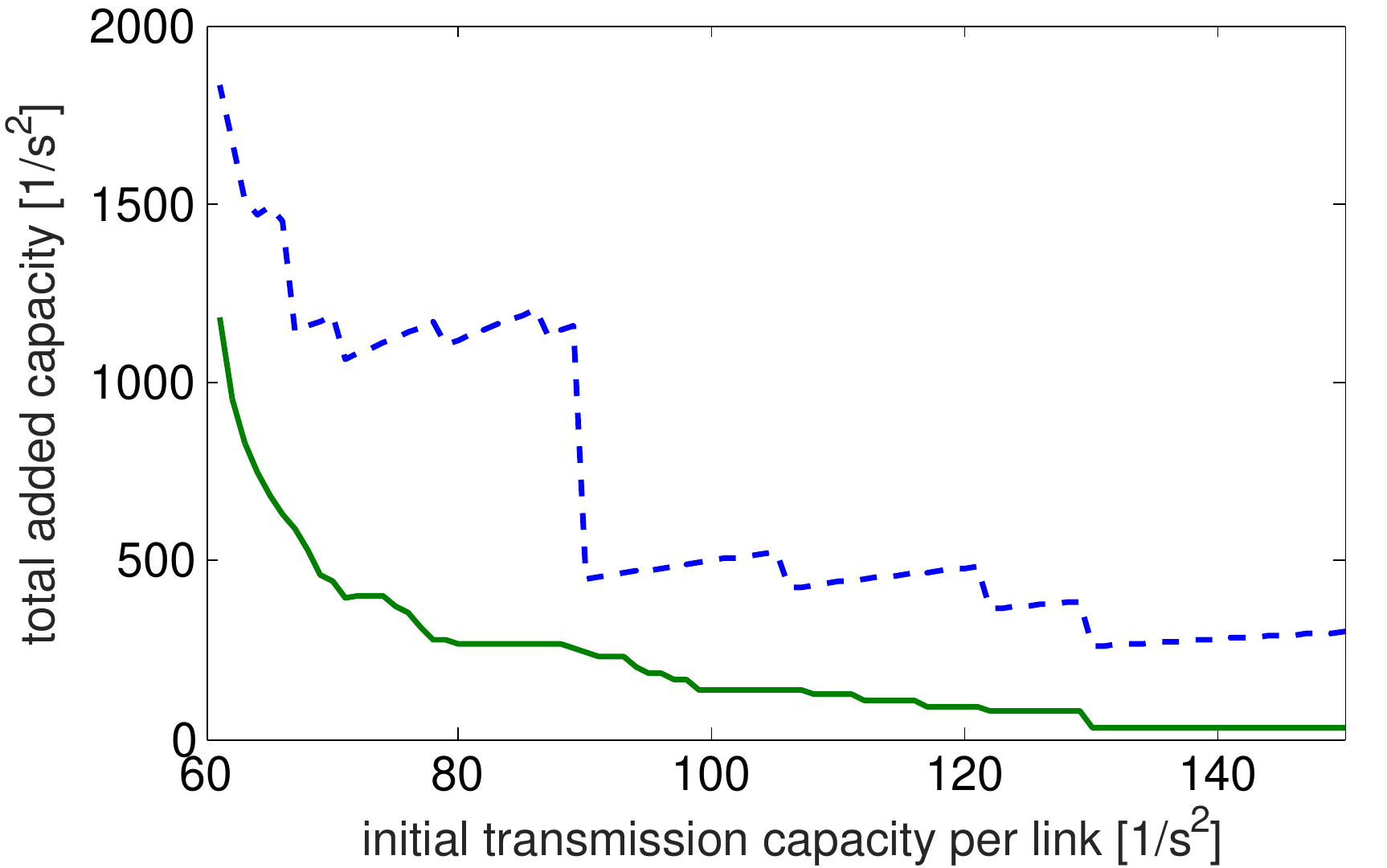}
\caption{\label{fig:figure5} (Color online) 
How much additional capacity is required to cure all critical links? Curing nonlocal bottlenecks (solid green, strategy I) requires less resources than building local backup lines (dashed blue, strategy II) for the test grid shown in Fig.~\ref{fig:figure1}. For further explanations see the text.}
\end{figure}

Our two strategies for improving network resilience are compared in Fig.~\ref{fig:figure5}. It shows the capacity that has to be added to cure all critical
links as a function of the initial transmission capacity assigned to the links. The steps in the dashed curve (strategy II) are due to the fact that this
strategy is directly sensitive to the number of critical links, which changes abruptly as a function of the capacity. The slight increase between succeeding
steps results from the fact that the added line for replacement of the critical link is equipped with the same capacity as the critical link was before its
removal. Adding capacities at the bottlenecks of the redundant capacities requires in all cases less additional capacity compared to building backup lines, so
strategy I is superior.

\subsection{Other network topologies}
One may have expected that a statement about the superior strategy can be given merely in terms of the network topology. In general, however, it is the
combination of the network topology with producers, consumers, line capacities and dynamics which decides about the superior strategy \cite{Hines10}. Keeping
everything unchanged, we illustrate for a regular ring topology with one shortcut (that is a small-world topology in the limit of an almost regular topology \cite{Watt98}) that either strategy I or strategy II is superior, depending on the capacity
assignment. Similarly, for specific choices of random networks we show that  which is the superior strategy also here depends on the gap of the capacities to the critical capacity value (below the critical capacity no stable state can be achieved).

For the small-world case we start from a ring of $N=100$ nodes with next-neighbor interactions and rewire just a single link. We randomly assign five large generators
with $P=10\, {\rm s}^{-2}$, ten small ones with $P=3.5\, {\rm s}^{-2}$ and 85 consumers with $P=-1\, {\rm s}^{-2}$. The capacities are homogeneously chosen for all
links. Now we rewire exactly one link, so that the ring is broken into a loop and an attached chain. For rewiring we choose a heavily loaded link, i.e., one that
is connected to a large power generator. We rewire the link such that the system can reach a stable state as long as the rewired link is present. The rewired link is
critical in the former sense.

\begin{figure}[tb]
\centering
\includegraphics[width=0.45\textwidth]{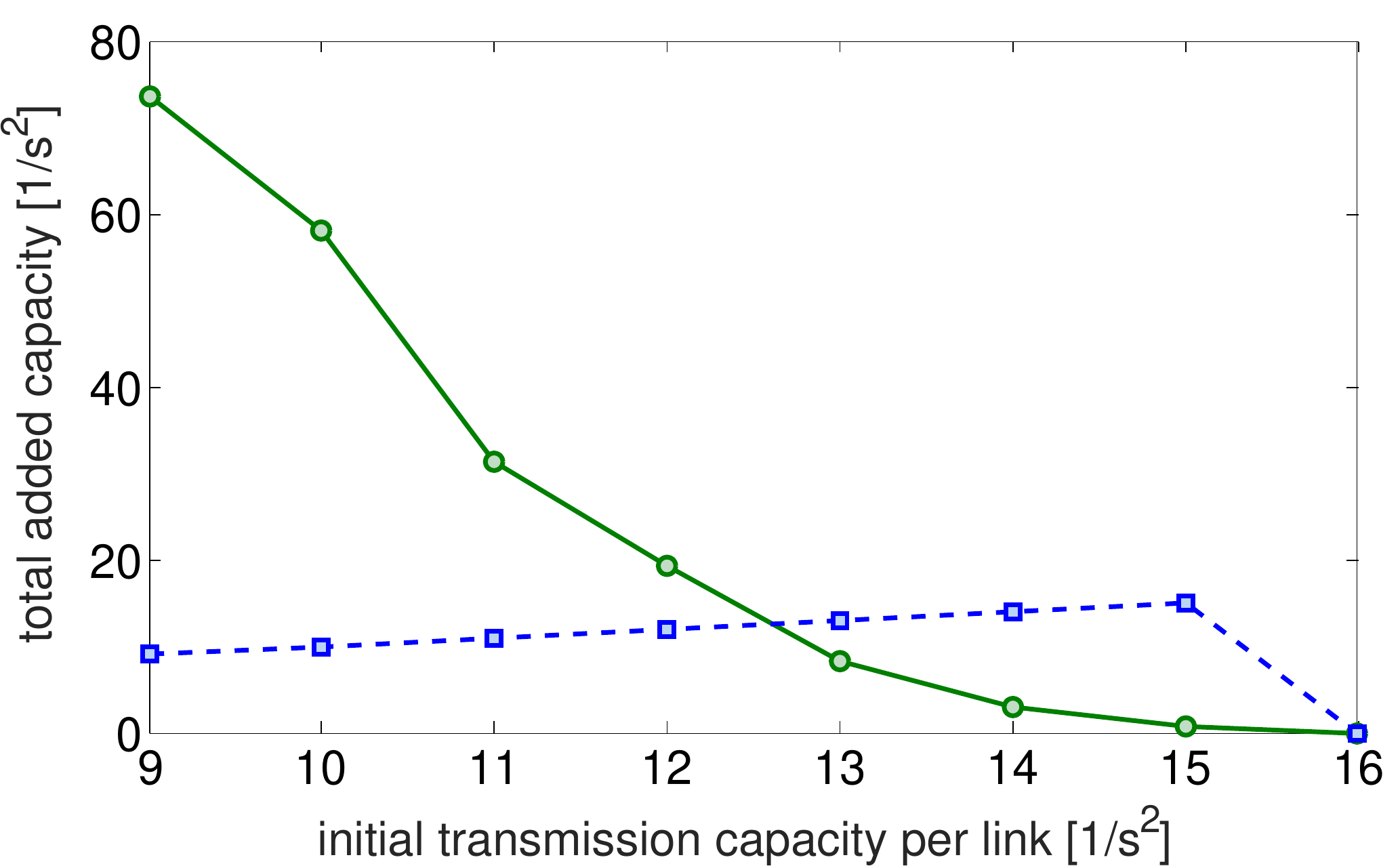}
\caption{\label{fig:smallworld} (Color online) Comparison of strategies as in Fig.~\ref{fig:figure5}, but for a regular ring topology with 100 nodes and a single shortcut. Depending on the assigned
capacities strategy I or strategy II is superior. The small-world grid consists of five large generators ($P=10\, {\rm s}^{-2}$), ten small generators
($P=3.5\, {\rm s}^{-2}$) and 85 consumers with $P=-1\, {\rm s}^{-2}$. Here the link from node 1 to 2 is rewired to node 50. For further details see the text.
}
\end{figure}

Depending on the assignment of capacities, strategy I or II may be superior, as demonstrated in Fig.~\ref{fig:smallworld}, where we determined the necessary additional
capacity to stabilize the rewired critical link. For high capacity values on each link strategy I is superior as it was the case for the Romanian power grid, while for
low capacity values strategy II is more efficient. We tested rewiring for other heavily loaded links with various destinations and found the same qualitative
behavior as indicated in Fig.~\ref{fig:smallworld}. In conclusion: For a fixed topology and fixed distributions of consumers and generators, the average load of the network is decisive. In highly loaded networks with little residual capacity strategy I requires too much additional capacity, so that strategy II is superior.

With the following example we present two random networks, which only differ by the density of links. We choose random networks of the Erd\"os Renyi
type \cite{Erdos1959}, again  with 100 nodes (five large generators with $P=10\, {\rm s}^{-2}$, ten small ones with $P=3.5\, {\rm s}^{-2}$ and 85 consumers
with $P=-1\, {\rm s}^{-2}$). The nodes are randomly chosen to be either one of the generators or one of the consumers. With probability $4/100$ or $6/100$, respectively,
we connect any randomly chosen pair of nodes, so that on average each node is connected with four or six other ones. The capacities are chosen homogeneously. For both connection probabilities, different realizations of the grid need different initial capacities $K_{{\rm min}}$ that are minimally required for reaching a
stable state. We then compare and average only over grids with similar $K_{{\rm min}}$. Here we consider a minimal transmission capacity of $K_{{\rm min}}=3.4\, {\rm s}^{-2}$ for an
average of four connections and of $K_{{\rm min}}=4.6\, {\rm s}^{-2}$ for six connections.

\begin{figure}[tb]
\centering
\includegraphics[width=0.45\textwidth]{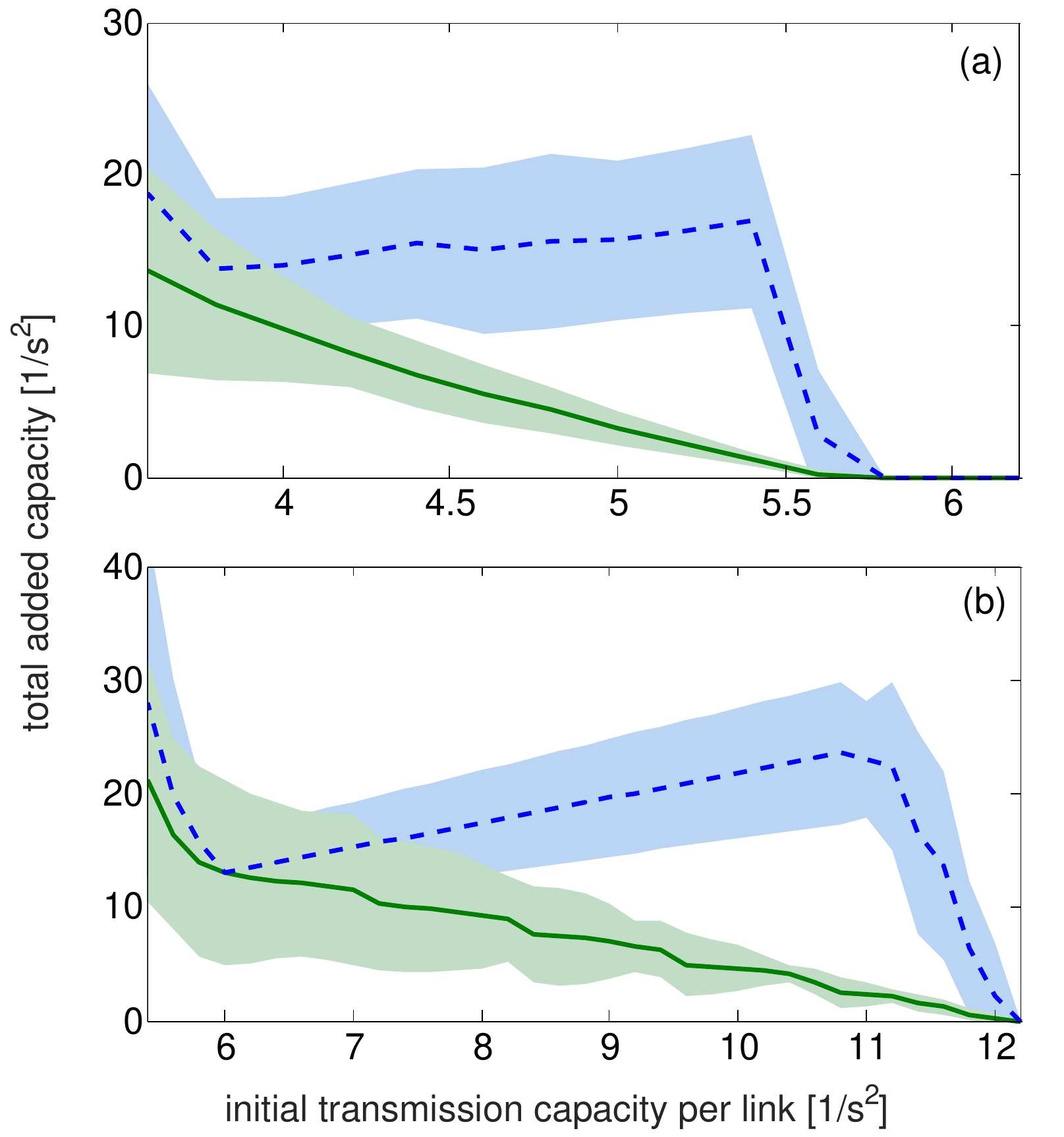}
\caption{\label{fig:random6}(Color online) Comparison of strategies as in Fig.~\ref{fig:figure5}, but for random networks with an average degree of six (a) and four (b). Depending on that, strategy I is almost always superior (a), or only for large initial transmission capacities (b). The
random grids consist of five large generators ($P=10\, {\rm s}^{-2}$), ten small generators ($P=3.5\, {\rm s}^{-2}$) and 85 consumers with $P=-1\, {\rm s}^{-2}$. The shaded areas indicate the standard deviation.
For further details see the text.
}
\end{figure}

The average and standard deviation of the total added capacity are displayed for both strategies I and II, where the average is taken over fifty realizations for an average degree of six (four)  connections between the nodes in Fig.~\ref{fig:random6}(a)((b)), respectively. In Fig.~\ref{fig:random6}(a) the
length of the shortest alternative paths varies between three and seven links, similar numbers as for the Romanian power grid. Here strategy I is almost always superior in the sense that
less additional capacity to cure all critical links is required than for strategy II. The only exception is for grids with low initial transmission capacities close to $K_{{\rm min}}$, not displayed here, similarly to the small-world case.

In Fig.~\ref{fig:random6}(b) with an average degree of four, the
length of the shortest alternative paths  varies between six and thirteen links. Here, strategy I  performs better outside the error bars for initial transmission
capacities, for which the gap to  $K_{{\rm min}}$ is even larger than in the previous case. This may be related to the fact that the length of the shortest alternative paths is in general longer for a less dense grid. With the length of the alternative paths also the probability increases to find several bottlenecks along them, whose capacities must be increased.

\section{Conclusions and Outlook}

Single link failures may induce global outages in power grids and other complex supply networks. The design of future grids which ensure $N-1$ security is generally based on large-scale numerical simulations. In this article we contribute to a better theoretical understanding of network resilience and optimization by comparing local and nonlocal strategies to extend existing grids.
Upgrading existing critical links seems a viable option, which, however, involves high costs (in terms of update capacity needed). We have introduced a novel curing strategy (strategy I) based on the mathematical analysis of flow rerouting  across redundant capacities on  the network. The method is intrinsically non-local, as it exploits properties of the entire collective dynamics of the network. It may provide a lower-cost solution compared to local link updating at the same collective impact of improving network robustness.

In principle redundancy in the grid provides an option for non-locally rerouting flows. We find that bottlenecks in these redundancies often cause outages induced by link failures. Actually the number of bottlenecks along the rerouting paths may be quite small (cf.~\cite{16critical}). Our strategy exploits this fact by systematically enhancing the capacities of these bottlenecks, thereby making use of the information also on the remote flow distribution.

This non-local strategy works well if the critical links are caused by a few pronounced bottlenecks rather than an overall heavily loaded grid. This in turn depends on the network topology as well as the distribution of generators (inputs) and consumers (outputs). For instance, non-local rerouting becomes hard or less advantageous than the local updating of a  critical link, if a large fraction of links is already heavily loaded in the original network and little non-local redundancy is available. This effect was seen for the regular ring topology and both versions of random networks. The effect is more pronounced if, in addition, the alternative paths between two nodes are long. This feature is specific to the topology. In our examples, the density of links  in the random grid with an average degree of four was lower than for degree six, leading to longer detours for the flow and possibly more bottlenecks along them.

Taken together, our results indicate that non-local rerouting of flows upon link failure often provides a viable complement to local updates. In a range of settings, the nonlocal strategy undercuts the costs of local updating. In power grids, taking into account non-local rerouting along the lines of our strategy I may enhance our options to satisfy the $(N-1)$-criterion in a more efficient way.

\ack

We gratefully acknowledge support from the Federal Ministry of Education and Research (BMBF grant no.~03SF0472A-E),  the Deutsche Forschungsgemeinschaft (DFG grant no. ME-1332/19-1 to M.R. and H.M-O.), the Helmholtz Association (via the joint initiative “Energy System 2050 – A Contribution of the Research Field Energy” and the grant no.~VH-NG-1025 to D.W.) and the Max Planck Society to M.T.

% --- Literatur -------------------------------------------------------------------

\end{document}